# Mouse Embedded Soft Vibration Actuator for Exploring Surface Textures

Asahi Kurokawa[1], Masaharu Shimizu[1], Mitsuhito Ando[1], and Haruo Noma[1]

[1] College of Information Science and Engineering, Ritsumeikan University, Osaka, Japan

(Email: akurokawa@mxdlab.net)

**Abstract ---** We aimed to develop a tactile display that allows users to actively explore the virtual texture of a surface. We developed a tactile display embedded in an optical mouse that provides a wide range of frequency vibrations to the user's fingertip in response to its movement. We conducted an experiment to confirm the degree of fineness based on a stripe pattern identification task using a trial system. The results showed that the subject could identify a fine spatial resolution up to 0.16 mm width.

## 1 INTRODUCTION

Online shopping has become popular; however, customers can only visually assess product quality through display images and descriptions. In physical stores, they can touch and feel products, enabling them to make purchase decisions based on tactile sensations, which are especially important for items such as clothing.

We feel the surface texture of products through various tactile senses including roughness, hardness, and warmth. To fully perceive surface textures, we must touch and trace the surface. Tracing a surface is a natural action that allows anyone to feel its texture. However, it is difficult to verify the tactility of a product when shopping online.

For a tactile display to be practical in everyday online shopping, it must be user friendly. Existing methods for tactile presentation include ultrasound [1] and electrical stimulations [2]. However, these methods often involve large devices or are challenging to wear, making them impractical for daily use. Another method involves the use of a rotating plate [3] that simulates the sensation of tracing by applying continuous shear force to the skin. However, the method can't effectively present textures beyond the material of the rotating plate.

To solve these problems, we aimed to allow users to feel the fine texture of a product displayed on a PC screen with a proposed tactile display. We focus on roughness as a key texture aspect and propose a tactile display that allows users to perceive roughness from an image. The proposed system enhances the perception of surface roughness by providing tactile feedback corresponding to the tracing motions of the user.

We have been developing a miniature actuator called the Soft Vibration Actuator (SVA) [4]. The SVA offers high controllability over a wide frequency range [5] and is sufficiently compact to integrate both the driving and presentation components. The goal was to enable tactile interaction with screen-displayed textures. In this paper, we reported the verification results of touch for fine textures.

## 2 SYSTEM IN RESPONSE TO TRACING MOTION

We developed the Mouse with Soft Vibration Actuator (SVA-M) to present vibrations in response to the user's movements on a flat surface. The SVA-M provides vibration at 120 Hz when the cursor is in the black area and stops vibrating in the white area on the PC screen. This method conveys information through vibration by tracing a black-and-white image. Fig. 1 shows a prototype of the SVA-M. The SVA[4] is an actuator that generates an electromagnetic force to vibrate liquid metal sealed in a soft tube. Fig. 2 shows a schematic of the principle of the SVA operation. Electromagnetic force is generated by passing an electric current through a liquid metal inside a soft tube. The soft tube oscillated depending on the switching frequency, which switched the direction of the current flow.

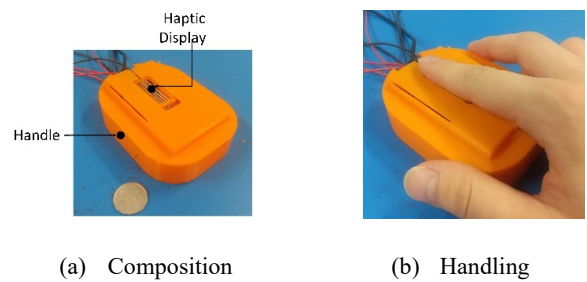

(a) Composition    (b) Handling

Fig.1  Soft Vibration Actuator embedded in Mouse (SVA-M)

In the SVA-M, we increased the number of tubes in the previous design. Four tubes were placed in parallel at 1 mm intervals. This design allows stimulation across the entire fingertip area and enhances tactile sensation. Fig. 3 shows the system diagram of SVA-M. An optical sensor (SPCP168A) detected mouse movements, and the color of the cursor's position on the PC screen determined the SVA-M vibration frequency. This allows the SVA-M to vibrate according to the motion of the user.

## 3 EVALUATION EXPERIMENT

### 3.1 Experimental Method

To evaluate this system, we experimented with six images of vertical stripes of varying line and space widths. Fig. 4 shows the six images. The line widths of these images were 1-pixel, 2-pixel, 4-pixel, 8-pixel, 16-pixel, and 32-pixel, respectively. With this setting, if the user moves the SVA-M by 40 mm in the real world, the cursor moves 1000 pixels on the screen. Table 1 shows the correspondence between the number of pixels on the screen and the real world distance for each image's linewidth. Henceforth, we define a 1-pixel image as the finest texture and a 32-pixel image as the coarsest texture. Participants in the experiment determined which of the two images was finer.

We instructed the participants to trace all samples as shown in Fig. 4 previously. During the experiment, the subjects did not watch a PC screen. The experimental procedure consisted of four steps. First, the participant touched the first texture for 5 seconds. Second, the participant rested for 1 s. Third, the participant touched the second texture for five seconds. Finally, the participant responds with which texture was finer. This procedure was performed for four sets, with each set consisting of 30 pairs. (120 trials in total). The order of texture presentation was randomized for each set and participant. We recorded the participants' responses and the coordinate data of the mouse cursor.

### 3.2 Experimental Environment

The PC screen used in this experiment had a 1920-pixel horizontal resolution, 1080-pixel vertical resolution, a 13.3-inch screen, and 60 Hz refresh rate. We employed

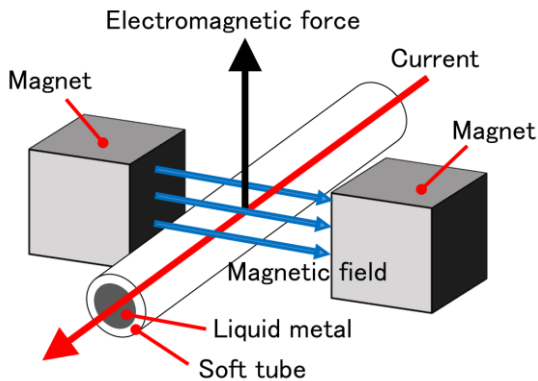

Fig.2 Principle of the SVA operation

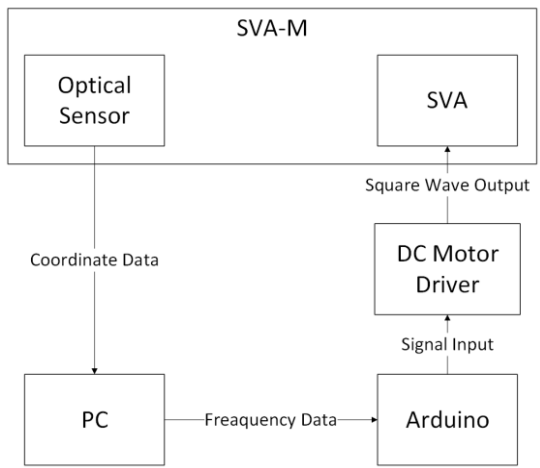

Fig.3 System diagram for SVA-M

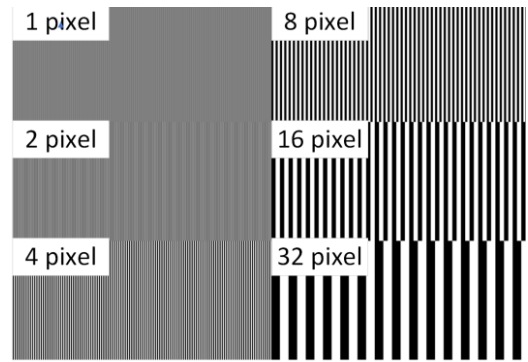

Fig.4 Images displayed in the experiment

Table 1  Length correspondence between screen and real world

| PC screen (pixel) | Real world (mm) |
|---|---|
| 1 | 0.04 |
| 2 | 0.08 |
| 4 | 0.16 |
| 8 | 0.32 |
| 16 | 0.64 |
| 32 | 1.28 |

Table 2 Percentage of trials answered as fine

| Pairs | | Texture that subjects answered as fine | | | | | |
|---|---|---|---|---|---|---|---|
| | | 1 pixel | 2 pixel | 4 pixel | 8 pixel | 16 pixel | 32 pixel |
| Comparison texture | 1 pixel | | 0.58 | 0.55 | 0.43 | 0.35 | 0.15 |
| | 2 pixel | 0.43 | | 0.50 | 0.53 | 0.33 | 0.13 |
| | 4 pixel | 0.45 | 0.50 | | 0.38 | 0.35 | 0.18 |
| | 8 pixel | 0.58 | 0.48 | 0.63 | | 0.38 | 0.13 |
| | 16 pixel | 0.65 | 0.68 | 0.65 | 0.63 | | 0.18 |
| | 32 pixel | 0.85 | 0.88 | 0.83 | 0.88 | 0.83 | |

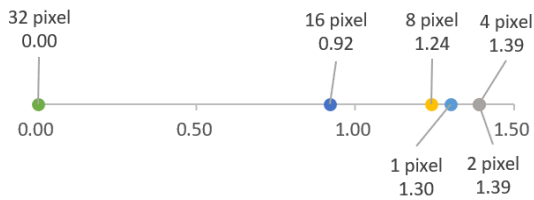

Fig.5 Scale value according to Thurstone's paired comparison method

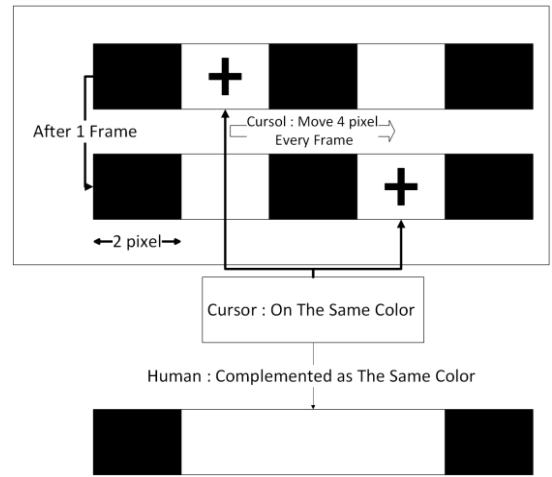

Fig.6 Scenes where textures are difficult to distinguish.

five male participants, aged 19-23 years, who had no visual or finger disabilities, as confirmed through a pre-test oral questionnaire

### 3.3 Results and Discussion

The generalizability of this study's findings may be limited due to age bias and the relatively small sample size. However, by conducting trials under consistent experimental conditions, a sufficient datas were collected. This allowed for the observation of certain trends in the results.

Table 2 shows the results of each pairwise comparison. The columns indicate the textures identified as finer and the rows indicate the compared textures. Each pair was tested 40 times. The values in the table show the percentage of trials in which the respondents judged the column texture to be finer. The percentages were rounded to two decimal places. All pairs containing the 32-pixel texture had a discrimination rate of 0.83 or higher, pairs containing the 16-pixel texture had a rate of 0.63 or higher, and most pairs containing the 8-pixel texture had a rate of 0.6 or higher. In contrast, combinations of 4-pixel, 2-pixel, and 1-pixel textures had a discrimination rate of 0.5 or less.

Based on the response rates, the fineness scale values were calculated using Thurstone's paired comparison method. Fig. 5 shows the calculated scale values. An internal consistency test was used to confirm the consistency of the values. This scale indicates the relative intensity of the fineness. For simplicity, the lowest value (32-pixel) is set to zero.

The scale values ordered the textures as 4-pixel, 2-pixel, 1-pixel, 8-pixel, 16-pixel, and 32-pixel, from highest to lowest fineness. For the 4-pixel, 8-pixel, 16-pixel, and 32-pixel textures, the perceived fineness order matched the linewidth thickness of the presented textures. The differences in scale values were 0.15 between 4-pixel and 8-pixel, 0.32 between 8-pixel and 16-pixel, and 0.92 between 16-pixel and 32-pixel. For the 1-pixel, 2-pixel, and 4-pixel textures, the order of the scale values was reversed compared to the actual texture order.

As shown in Fig. 5 that for 4-pixel, 8-pixel, 16-pixel, and 32-pixel textures, the perceived fineness scale order matches the actual line width order. The difference between the 4-pixel and 8-pixel textures was 0.15, which is a small but significant difference compared to coarser textures. Thus, for line widths between 0.16 mm and 1.28 mm, the perceived difference in fineness is influenced by the actual texture fineness.

However, the perceived order was reversed for the 1-pixel, 2-pixel, and 4-pixel line widths, with discrimination rates below 0.5. The average cursor movement speed was approximately 240 pixel/s for these combinations. We consider that the low discrimination rate for finer textures was due to the cursor skipping certain pixels. These results are explained by the example shown in Fig. 6. In this case, the cursor moved four pixels per frame at a display refresh rate of 60 Hz. It remained on the same color for all periods. This possibly caused improper vibration switching.

The evaluation of texture fineness in this experiment has several limitations. The study focused on textures that could be distinguished within a short timeframe.It is possible that finer textures could be identified by giving participants more time for tracing. Additionally, increasing the display's refresh rate could help suppress aliasing effects, potentially enhancing the accuracy of texture discrimination.

These results indicate distinguishing between textures with a fineness of 0.16 mm or less is difficult. However, differences are perceivable for textures coarser than 0.16 mm. We obtained this result because we had limited visual information. During real world exploratory activities involving surface textures, people use visual and tactile information. When people perceive a difference between visual and tactile fineness, they adjust the mouse movement speed. Therefore, we should focus on developing a system that simultaneously presents visual and tactile information in future work.

## 4 Conclusion

In this study, we developed SVA-M, which allows users to feel product surface textures. We conducted experiments using vertical stripe patterns to evaluate the texture fineness. The results indicate that the SVA-M can perceive textures with a fineness greater than 0.16 mm.

Future work will focus on presenting complex textures and multimodal sensations. Currently, the system handles only simple textures and cannot present the surface textures of real objects. To achieve a more detailed texture presentation, we plan to develop a system that controls the frequency of the SVA according to the speed and position of the virtual texture. Additionally, by visually displaying these textures, we aim to provide a more realistic experience of exploring virtual textures.

### Acknowledgement

This work was supported by JSPS KAKENHI Grant Number 22H00542.